\documentstyle[aps,prl,twocolumn]{revtex}
\tighten
\begin{document}
\draft
\title{CERN-TH/2002-193  \\
~~~ \\
~~~ \\
Cosmological Sign of Neutrino CP Violation}
\author
{P.H. Frampton$^{(1,2\dagger)}$, S.L. Glashow$^{(3)}$
and T. Yanagida$^{(1,4\dagger)}$}
\address
{(1) TH Division, CERN, CH 1211 Geneve, 23, Switzerland }
\address
{(2) Department of Physics and Astronomy,
University of North Carolina, Chapel Hill, NC27599, USA. }
\address{(3) Department of Physics, Boston University, Boston, MA 02215, USA.}
\address{(4) Department of Physics, University of Tokyo, Tokyo 113-0033, Japan.}
\address{$\dagger$ permanent address}

\maketitle

\begin{abstract}
It is shown how, in a class of models, the sign of the baryon number
of the universe can be related to CP violation
in neutrino oscillation experiments.
\end{abstract}

\medskip
\bigskip
\medskip

One of the most profound ideas is\cite{Sakharov} 
that baryon number asymmetry arises in the early universe
because of processes which violate CP symmetry and that 
terrestrial experiments on CP violation
could therefore inform us of the details of such 
cosmological baryogenesis.

The early discussions of baryogenesis focused on
the violation of baryon number and its possible relation to
proton decay. In the light of present evidence for neutrino masses
and oscillations 
it is more fruitful to associate the baryon number
of the universe with violation of lepton number\cite{FY}. 
In the present Letter
we shall show how, in one class of models, the sign of the
baryon number of the universe correlates with
the results of CP violation in neutrino oscillation
experiments which will be performed in the forseeable
future.

Present data on atmospheric and solar neutrinos suggest
that there are respective squared mass differences
$\Delta_a \simeq 3 \times 10^{-3} eV^2$
and
$\Delta_s \simeq 5 \times 10^{-5} eV^2$.
The corresponding mixing angles $\theta_1$
and $\theta_3$ satisfy
$tan^2 \theta_1 \simeq 1$
and $0.6 \leq sin^2 2\theta_3 \leq 0.96$
with $sin^2 \theta_3 = 0.8$ as the best fit.
The third mixing angle is much smaller than the other
two, since the data require $sin^2 2 \theta_2 \leq 0.1$.

A first requirement is that our model accommodate
these experimental facts at low energy.

In the minimal standard model, neutrinos are massless.
The most economical addition to the standard model
which accommodates both neutrino masses
and allows the violation
of lepton number to underly the cosmological baryon asymmetry
is two right-handed neutrinos $N_{1,2}$. 
This gives rise to a more constrained see-saw mechanism
than envisioned in \cite{Y}.

These lead to new terms in the lagrangian:

\begin{eqnarray}
{\cal L} & = & \frac{1}{2} (N_1, N_2) \left(
\begin{array}{cc} M_1 & 0 \\ 0 & M_2
\end{array} \right)
\left( \begin{array}{c} N_1 \\ N_2 \end{array} \right)
+  \nonumber \\
& + & (N_1, N_2)
\left( \begin{array}{ccc} a  &  a^{'}  &  0  \\
0  &  b  &  b^{'}  \end{array} \right)
\left( \begin{array}{c} l_1  \\  l_2 \\ l_3 \end{array}
\right)  H  + h.c.
\label{Lag}
\end{eqnarray}
where we shall denote the rectangular Dirac mass matrix
by $D_{ij}$. We have assumed a texture
for $D_{ij}$ in which the upper
right and lower left entries vanish.
The remaining parameters in our model
are both necessary and sufficient
to account for the data.
The texture zeroes of Eq(\ref{Lag}) can be achieved with supersymmetry
by imposing a complicated global symmetry.

For the light neutrinos, the see-saw mechanism leads to
the mass matrix\cite{Y}
\begin{eqnarray}
\hat{L} & = & D^T M^{-1} D \nonumber \\
& = & \left( \begin{array}{ccc}
\frac{a^2}{M_1}  &  \frac{a a^{'}}{M_1}  &  0  \\
\frac{a a^{'}}{M_1}  &  \frac{(a^{'})^2}{M_1} + \frac{b^2}{M_2}  
&  \frac{b b^{'}}{M_2} \\
0  &  \frac{b b^{'}}{M_2}  &  \frac{(b^{'})^2}{M_2}  
\end{array}  \right)
\label{L}
\end{eqnarray} 

We take a basis where $a, b, b^{'}$ are real and where
$a^{'}$ is complex $a^{'} \equiv |a^{'}|e^{i \delta}$.
To check consistency with low-energy
phenomenology we temporarily take the specific
values (these will be loosened later)
$b^{'} = b$ and $a^{'} = \sqrt{2} a$ and all parameters real.
In that case:
\begin{eqnarray}
\hat{L} 
& = & \left( \begin{array}{ccc}
\frac{a^2}{M_1}  &  \frac{\sqrt{2}a^2}{M_1}  &  0  \\
\frac{\sqrt{2}a^2}{M_1}  &  \frac{2a^2}{M_1} + \frac{b^2}{M_2}  
&  \frac{b^{2}}{M_2} \\
0  &  \frac{b^{2}}{M_2}  &  \frac{b^{^2}}{M_2}  
\end{array}  \right)
\label{LL}
\end{eqnarray} 
We now diagonalize to the mass basis by writing:
\begin{equation}
{\cal L} = \frac{1}{2} \nu^T \hat{L} \nu 
= \frac{1}{2} \nu^{'T} U^T \hat{L} U \nu^{'}
\end{equation}
where
\begin{eqnarray}
U & = & \left( \begin{array}{ccc} 1/\sqrt{2}  &  1/\sqrt{2}  &  0  \\
- 1/2 & 1/2  &  1/\sqrt{2}  \\
1/2  &  -1/2 &  1/\sqrt{2}
\end{array} \right) \times \nonumber \\
& \times &  
\left( \begin{array}{ccc}
1  &  0  &  0  \\
0  &  cos\theta &  sin\theta  \\
0  &  - sin\theta  &  cos\theta  
\end{array}
\right)
\end{eqnarray}
We deduce that the mass eigenvalues  and $\theta$
are given by

\begin{equation}
m(\nu_3^{'}) \simeq 2 b^2/M_2; ~~~ m(\nu_2^{'}) \simeq 2 a^2 /M_1; ~~~ 
m(\nu_1^{'}) = 0
\end{equation}
and
\begin{equation}
\theta \simeq m(\nu_2^{'}) / (\sqrt{2} m(\nu_3^{'}))
\end{equation}
in which it was assumed that $a^2/M_1 \ll b^2/M_2$.

By examining the relation between the three mass eigenstates
and the corresponding flavor eigenstates
we find
that for the unitary matrix relevant to neutrino oscillations
that
\begin{equation}
U_{e3} \simeq sin\theta/\sqrt{2} \simeq m(\nu_2)/(2m(\nu_3))
\end{equation}

Thus the assumptions $a^{'} \ \sqrt{2} a$, $
b^{'} = b$ adequately fit the
experimental data, but
$a^{'}$ and $b^{'}$ could 
be varied around
$\sqrt{2}a$ and $b$ respectively
to achieve better fits.

But we may conclude that
\begin{eqnarray}
2b^2/M_2 & \simeq & 0.05 eV = \sqrt{\Delta_a} \nonumber \\ 
2a^2/M_1 & \simeq & 7 \times 10^{-3} eV  = \sqrt{\Delta_{s}}
\label{numass}
\end{eqnarray}

\noindent It follows from these values,
taking into account that the $L$ asymmetry for $N_2$
decay is suppressed due to the washing-out effect,
also by a factor $(hh^{\dagger})^2/(hh^{\dagger})_{22}$,
that the mass
$M_2$ for $N_2$ must be larger than $10^{12}$ GeV\cite{buchpascos}
if $N_2$ decay is responsible for the present baryon asymmetry.

This would mean that the reheating temperature is 
higher than $10^{12}$ GeV which would lead to a serious
problem with cosmological production of gravitinos of mass O(TeV)
in the supersymmetric standard model.    

Therefore we consider it more likely that $N_1$ decay 
produces the baryon asymmetry, since the mass
$M_1$ of $N_1$ can be $\sim 10^{10}$ GeV.

Making this choice
enables us to 
compute from the sign (known from cosmological B)
of the high-energy
CP violating parameter ($\xi_H$) appearing
in leptogenesis the sign of the CP violation parameter
which will be measured in low-energy
$\nu$ oscillations ($\xi_L$).

We find the baryon number $B$ of the universe
produced by $N_1$ decay
proportional to\cite{Buch}
\begin{eqnarray}
B & \propto & \xi_H = 
(Im D D^{\dagger} )_{12}^2 = Im (a^{'} b)^2 \nonumber \\
& = & + Y^2a^2b^2 sin 2\delta
\label{highenergy}
\end{eqnarray}
in which $B$ is positive by observation of the universe.
Here we have loosened our assumption about $a^{'}$
to $a^{'} = Y a e^{i \delta}$.

At low energy the CP violation in neutrino oscillations is governed by
the quantity\cite{branco}
\begin{equation}
\xi_L = Im (h_{12} h_{23} h_{31})
\end{equation}
where $h = \hat{L} \hat{L}^{\dagger}$.

Using Eq.(\ref{L}) we find:
\begin{eqnarray}
h_{12} & = & \left( \frac{a^3 a^{'}}{M_1^2} + \frac{a |a^{'}|^2 a^{'*}}{M_1^2}
\right) + \frac{a a^{'} b^2}{M_1M_2} \nonumber \\
h_{23} & = & \left( \frac{b b^{'} a^{'2}}{M_1 M_2} \right)
 + \left( \frac{b^3 b^{'}}{M_2^2} + \frac{b b^{'3}}{M_2^2} \right)
\nonumber \\
h_{31} & = & \left( \frac{a a^{'*} b b^{'}}{M_1 M_2} 
\right) \nonumber \\
\end{eqnarray}
from which it follows that
\begin{equation}
\xi_L =  - \frac{a^6 b^6}{M_1^3 M_2^3} sin 2\delta [ Y^2 (2 + Y^2)]
\label{lowenergy}
\end{equation}
Here we have taken $b=b^{'}$ because
the mixing for the atmospheric neutrinos
is almost maximal.

Neutrinoless double beta decay $(\beta\beta)_{0\nu}$
is predicted at a rate corresponding to $\hat{L}_{ee} \simeq 3 \times 10^{-3}eV$.

The comparison between Eq.(\ref{highenergy})
and Eq.(\ref{lowenergy})
now gives a unique relation between the signs of $\xi_L$ and $\xi_H$.

As a check of this assertion we consider
the equally viable alternative model 

\begin{equation}
D = \left( \begin{array}{ccc} a & 0 & a^{'} \\
0 & b & b^{'}
\end{array} \right)
\label{alternative}
\end{equation}
in Eq.(\ref{Lag}) 
where $\xi_L$ reverses sign but the
signs of $\xi_H$ and $\xi_L$ are
still uniquely correlated once the $\hat{L}$
textures arising from the $D$ textures
of Eq.(\ref{Lag}) and Eq.(\ref{alternative})
are distinguished by low-energy phenomenology.
Note that such models have five parameters including
a phase
and that cases B1 and B2 in \cite{FGM}
are unphysical limits of (\ref{Lag})
and (\ref{alternative}) respectively.

This fulfils in such a class of models the idea of \cite{Sakharov}
with only the small change that baryon
number violation is replaced by lepton number violation.

\bigskip

The work of P.H.F. was supported in part
by the Department of Energy
under Grant Number
DE-FG02-97ER-410236;
that of S.L.G.
by the National Science Foundation
under Grant Number
NSF-PHY-0099529;
and that of T.Y.
by Grant-in-Aid for Scientific Research (S)14102004.

\end{document}